\documentclass[fleqn,usenatbib]{mnras}

\usepackage{newtxtext,newtxmath}
\usepackage{graphicx}
\usepackage{booktabs}
\usepackage{epstopdf}
\usepackage{rotating}
\DeclareGraphicsExtensions{.png,.pdf}
\usepackage[T1]{fontenc}

\DeclareRobustCommand{\VAN}[3]{#2}
\let\VANthebibliography\thebibliography
\def\thebibliography{\DeclareRobustCommand{\VAN}[3]{##3}\VANthebibliography}

\usepackage{graphicx}	
\usepackage{amsmath}	

\usepackage{natbib}
\usepackage{subfig}
\usepackage{booktabs}
\usepackage{upgreek}
\usepackage{setspace}
\usepackage{gensymb}
\usepackage{comment} 
\usepackage{multirow}
\usepackage{changepage}
\usepackage{float}
\usepackage{appendix}
\usepackage{tabularx}
\usepackage{longtable}
\usepackage{epsfig}
\usepackage{morefloats}
\usepackage{pdflscape}
\usepackage{fancyhdr} 

\fancypagestyle{mylandscape}{
\fancyhf{} 
\fancyfoot{
\makebox[\textwidth][r]{
  \rlap{\hspace{.75cm}
    \smash{
      \raisebox{4.87in}{
        \rotatebox{90}{\thepage}}}}}}
}

\newcommand{\eal}[2]{\ifmmode{\mathrm{#1\,#2}}\else{#1\textsc{$\,$\lowercase{#2}}}\fi\xspace}
\newcommand{\feal}[2]{\ifmmode{\mathrm{#1\,#2}}\else{[#1\textsc{$\,$\lowercase{#2}}]}\fi\xspace}
\newcommand{\hfeal}[2]{\ifmmode{\mathrm{#1\,#2}}\else{#1\textsc{$\,$\lowercase{#2}}]}\fi\xspace}

\title[Shock shaping of Nova V906 Car]{Shock shaping? Nebular Spectroscopy of Nova V906 Carinae}

\author[É. J. Harvey et al.]{
É. J. Harvey,$^{1}$\thanks{E-mail: eamonnharvey@gmail.com}
E. Aydi,$^{2}$
L. Izzo,$^{3}$
C. Morisset,$^{4}$
M. J. Darnley,$^{1}$
K. Fitzgerald,$^{5}$
P. Molaro,$^{6}$
\newauthor
F. Murphy-Glaysher,$^{1}$
M. P. Redman$^{7}$
and M. Shrestha,$^{1,8}$
\\
$^{1}$ Astrophysics Research Institute, Liverpool John Moores University, IC2 Liverpool Science Park, Liverpool, L3 5RF, UK \\
$^{2}$Center for Data Intensive and Time Domain Astronomy, Department of Physics and Astronomy, Michigan State University, East Lansing, MI 48824, USA \\
$^{3}$DARK, Niels Bohr Institute, University of Copenhagen, Jagtvej 128, 2200 Copenhagen \\
$^{4}$Universidad Nacional Autónoma de México, Instituto de Astronomía, AP 106,  Ensenada 22800, BC, México \\
$^{5}$Technological University of the Shannon, Athlone Campus, Co. Westmeath, N37HD68, Ireland \\
$^{6}$Osservatorio Astronomico di Trieste, Vuz GB Trepolo 11 34131, Trieste, Italy \\
$^{7}$Centre for Astronomy, National University of Ireland Galway, Galway, Ireland \\
$^{8}$ Steward Observatory, University of Arizona, 933 North Cherry Avenue, Tucson, AZ 85721-0065, USA\\
}

\date{Accepted XXX. Received YYY; in original form ZZZ}

\pubyear{2023}

\begin{document}
\label{firstpage}
\pagerange{\pageref{firstpage}--\pageref{lastpage}}
\maketitle

\begin{abstract}
V906 Carinae was one of the best observed novae of recent times.
It was a prolific dust producer and harboured shocks in the early evolving ejecta outflow. 
Here, we take a close look at the consequences of these early interactions through study of high-resolution UVES spectroscopy of the nebular stage and extrapolate backwards to investigate how the final structure may have formed. A study of ejecta geometry and shaping history of the structure of the shell is undertaken following a spectral line $\textsc{SHAPE}$ model fit. A search for spectral tracers of shocks in the nova ejecta is undertaken and an analysis of the ionised environment. Temperature, density and abundance analyses of the evolving nova shell are presented.

\end{abstract}




\section{Introduction}

Classical nova events are characterised by observed increases in visual brightness of 6-12 magnitudes and are due to eruptions on the surface of a white dwarf in a binary system. Much has been learned from the optical photometric evolution of novae and they are normally classified by their optical eruption spectra and light curves. The photometric decline time of a nova is characterised by its $t_{2}$ or $t_{3}$, the time taken to decline by 2 or 3 magnitudes, respectively, from maximum optical light. \cite{93lc} classify 
a variety of nova eruption light curves and give physical explanations for many of their features. Many of these light curves can be characterised simply by a fast rise and a smooth decline. However, they can often feature transitional features such as plateaus, flat-tops, cusps, jitters, plus other fast and slow temporary increases in brightness during the nova decline phase.

Spectroscopic observations, however, are usually required to distinguish novae from other transients as well as classifying them within established nova taxonomy. For this reason, novae have been observed spectroscopically since T Aur \citep[1891; see][]{payne}, and have been studied systematically since \cite{1991ApJ...376..721W}. A commonly adopted classification scheme for nova spectroscopy is that known as the Tololo scheme, first presented in \citep{1991ApJ...376..721W,tololo}. Nova spectra are characterised by several observable stages during and post eruption. The spectral fingerprint of a nova may be derived either if the nova shows some critical features, or else if the nova does not show certain stages. However, a new paradigm now exists following the detection of novae in the $\gamma$-ray regime \citep{gammanov} as well as the growing availability of high-cadence, medium/high-resolution spectroscopy from large ground based telescopes \citep{2020ApJ...905...62A}. Novae have also been recently unveiled as one of the main lithium factories in the universe \citep{V906Car_4_7BE_2020MNRAS.492.4975M}.

The shaping mechanisms of nova shells are probes of the processes that take place at energy scales between 
planetary nebulae and supernova remnants. 
However, the three-dimensional structure of nova shells is difficult to untangle when viewed on the plane of the sky. This is true even when spatially resolved imaging of the nova shell is available, which is not the case for recent nova events. In order to pin down the geometry, one must use the observations that are commonly accessible to them. In the optical domain this is mostly photometric observations and a growing wealth of spectroscopic observations. Although, polarimetric and spectro-polarimetric observations are increasingly being gathered for their diagnostic potential on system orientation and scattering properties \citep{evanspol,me_V5668Sgr,kawahita}.

V906 Carinae (2018) was a bright nova that was observed across the electromagnetic spectrum. It has even served as a basis for the development of a classical nova unification theory \citep{2020ApJ...905...62A}. V906 Car was discovered on the rise (UT 2018-03-16.32), saturating the 14cm Cassius telescope detector, reported in \cite{V906Car_discovery_2018ATel11454....1S}. As it was discovered by the ASASSN network in 2018 it was named ASASSN-18fv \citep{2014AAS...22323603S}. Long baseline optical and NIR photometery are provided in \cite{V906Car_photometry_2020ApJ...899..162W} of the DQ Her-like light curve, where the extinction towards the nova as well as distance are derived. \cite{V906Car_photometry_2020ApJ...899..162W} report  $M_\mathrm{WD} < 0.8\,\mathrm{M}_{\odot}$ (or $0.71\,\mathrm{M}_{\odot}$, from modelling the accretion disk of the quiescent system) and a companion star possibly of G or K type with mass $0.23-0.43\,\mathrm{M}_{\odot}$. $t_{2}$ is difficult to define for novae with non-smooth declines. The nova was not detected in the radio regime \cite{V906Car_radio_2018ATel11504}. The X-ray characteristic behaviour is summarised in \cite{V906Car_xray_2020MNRAS.497.2569S}, where solutions to column densities, temperatures and elemental abundances are presented. \cite{2020MNRAS.495.2075P} find an ejected mass of 
$6\times10^{-4}\,\mathrm{M}_{\odot}$ from a detailed photoionisation analysis of the ejecta. 

V906 Car is the brightest $\gamma$-ray detected nova to date \citep{V906Car_gamma_2018ATel11553....1P,V906Car_Aydi_Nature_2020NatAs...4..776A}. With the $\gamma$-ray emission being associated with shocks internal to the ejecta. The shocks are hypothesised to originate from the interaction of an initial slower, equatorial bound, denser ejection and subsequent higher velocity, but less massive ejection events. The presence of faster subsequent ejections (or high-velocity winds of finite duration) are seen by the appearance of high velocity features in optical spectroscopy shown in \cite{2020ApJ...905...62A}. 
The appearance of high velocity features in the spectra appear coincident with short-duration flare episodes in the early light curve. The interaction of the ejection episodes then appear to coincide with longer-duration optical rebrightenings. 
This is affirmed through the analysis of the concurrence of the optical and $\gamma$-ray emission peaks, albeit only available at later times due to Fermi-LAT down-time.  

The geometrical structure of the expelled shell from the V906 Car nova has been studied primarily in \cite{2020MNRAS.495.2075P}, where they fit an evolving morpho-kinematic model using the $\textsc{SHAPE}$ code \citep{shape} over several epochs. They find an inclination of the system of 53$^{\textrm{o}}$ and arrive at a slightly asymmetric nebular morphology. The spectral lines used for modelling are not yet in their frozen state, although are primarily isolated lines. This is suspected to lead conclusions towards an asymmetric structure, at odds with the population of well resolved classical nova shells that are found to be axisymmetric structures. Before nova shell lines have entered their frozen state they are subject to internal reddening effects in an optically thick shell.  \cite{2020MNRAS.495.2075P} give solutions on the position angle of the shell, however, we hold the position angle of the nova shell constant as the object is much less extended than the slit width at time of observation, and do not have access to polarimetric data to inform the structure's position angle. With this said, the shell inclination and structural components found in \cite{2020MNRAS.495.2075P} are similar to those found in this work. Also, the results of the $\textsc{Cloudy}$ simulations conducted in \cite{2020MNRAS.495.2075P} are found to complement our own, considering the base models are for different epochs. In the work \cite{2021MNRAS.505.2518M} the geometry of the outflow of V906 Car is also discussed. The interpretation of \cite{2021MNRAS.505.2518M} suggests the spectral line signatures to be related to a slow moving spherical ejection, jets and an accretion disk. The jet paradigm relies on the accretion disk being, at least, partially responsible for the launching of a jet, such that the accretion disk must re-establish before the jet is launched. A theoretical model and explanation of the launching of gamma-rays within this model is assumed to be in progress. Considering the shaping mechanism of nova shells to be still open question, jets may have a role. 

The origin of transient heavy element absorption (THEA) lines in nova spectra has proven to be a point of contention in the literature. They were first proposed by \cite{2008ApJ...685..451W} to originate in a circumbinary disk, fed by an outflow from the companion star. Adding to this 
\cite{V906Car_circumFeO_2020MNRAS.494..743M} arrive at the conclusion of the existence of circumbinary Fe and O for the V906 Car system and claim this reservoir may provide the origin of the THEA lines. The argument for the spectral features not originating in the shell itself is from the lower velocities observed. However, in \cite{2020ApJ...905...62A} it is suggested that the THEA lines originate instead in the ejecta, the dense (equatorial dominated) outflow expelled from the white dwarf i.e., the early, slower component. The evolution of the THEA lines were found to have the same structure as Balmer, Fe and other lines associated with the expelled nova shell. Also, in this work we find the $\lq$M' spectral line shape to be representative of an unblended shell line at the corresponding epoch to \cite{V906Car_circumFeO_2020MNRAS.494..743M}.  

The reader is recommended to refer to \cite{V906Car_Aydi_Nature_2020NatAs...4..776A,2020ApJ...905...62A} for the evolution of the high velocity features observed early on in the nova and to \cite{2020MNRAS.495.2075P} for a classical treatment of the pre-nebular phase spectral evolution.


\section{Observations}

\subsection{VLT-UVES optical spectroscopy.}
This publication is based on a series of spectra of V906 Car obtained with the Ultraviolet and Visual Echelle Spectrograph (UVES) \citep{2000SPIE.4008..534D} mounted at the UT2 telescope of the European Southern Observatory (ESO) in Paranal, Chile (Obs ID.\ 0100.D-0621 PI: Paolo Molaro; Obs ID.\ 2100.D-5048 PI: Paolo Molaro; Obs ID.\ 0103.D-0764, PI: Luca Izzo). Out of a set of 20 obtained spectra, concentrated during the early evolving epochs from 2018 March 20.65 to 2018 July 5.0 \citep[that have been covered extensively in][]{V906Car_Aydi_Nature_2020NatAs...4..776A,V906Car_4_7BE_2020MNRAS.492.4975M,2020ApJ...905...62A} we instead focus here on the later spectrum from 2019 March 22.1 (day +395 post-maximum (used in this paper), or day 405 post-eruption as used in \cite{V906Car_Aydi_Nature_2020NatAs...4..776A,2020ApJ...905...62A}), covering the nebular phase of the nova. The instrumental configuration for all the epochs was the DIC1 346-564,  with central  wavelength of 346 nm (range 305--388 nm) in the blue arm and  564 nm (460--665 nm) in the red arm. Every observation was followed by another with setting  DIC2 437-760, which cover the 380--490 nm region and the 700--1000 nm region. The log of early UVES observations is reported in Table. D.1 of \cite{2020ApJ...905...62A}  and available at  \url{http://www.astrosurf.com/aras/Aras_
DataBase/Novae/2018_NovaCar2018.htm}. A full list of observations used in this work can be seen in Table. \ref{tab:UVESobs}. The observations were reduced using the reflex environment \cite{2013A&A...559A..96F}.

\begin{table}
\caption{Diary of the UVES observations used for this work. The slit width was 0.6" for all observations using filter Her 5. A wavelength range of 3050 to 9000 \AA.}\label{tab:UVESobs}
\begin{center}
\scriptsize
\begin{tabular}{rrrrrr}
\hline
\hline
\multicolumn{1}{c}{{MJD}} & 
\multicolumn{1}{c}{{Day}} & 
\multicolumn{1}{c}{exp (s)}& 
\multicolumn{1}{c}{R/1000}& 
\multicolumn{1}{c}{exp (s)}&
\multicolumn{1}{c}{R/1000}\\
\multicolumn{1}{c}{} & 
\multicolumn{1}{c}{a.m.}&
\multicolumn{1}{c}{346 nm} &
\multicolumn{1}{c}{} & 
\multicolumn{1}{c}{564 nm} &
\multicolumn{1}{c}{}  \\
   \hline
    58199.1438& -6  & 60 & 59    & 60&66   \\
    58199.1452&   & 300&59     & 300&66   \\
    58201.1412& -4  & 300&59     & 60&66   \\
    58203.0924& -2  & 300&59     & 60&66   \\
    58205.0687& 0  & 300&59     & 60&66   \\
    58207.0599& 2  & 600&59     & 120&66   \\
    58209.0218& 4  & 600&59     & 120&66   \\
    58213.0665& 8  &  600&59     & 120&66   \\
    58215.1292& 10  &  600&59     & 120&66   \\
    58217.0686& 12  &  600&59     & 120&66   \\
    58220.1292& 15  &  600&59     & 120&66   \\
    58223.1262& 18  &  600&59     & 60&66   \\

    58228.0002& 22  & 150&59    & 30&66   \\

    58235.9959& 30  &  300&59     & 60&66   \\
 
    58303.9851& 98  & 1400&59     & 13x60&66   \\

    58598.0729& 395  &  460&59     & 90&66  \\
        \hline
        \hline
\end{tabular}
\end{center}
\end{table}

\subsection{Goodman Spectrograph on SOAR}

A spectrum was obtained on 12 December 2020 at 07:55:10 UTC in order to view the shell in its pure nebular state, this corresponds to the +1001 post eruption epoch. The spectrum was obtained using the Goodman spectrograph mounted on the 4.1 m SOAR telescope. The spectrograph was used in its red configuration, with a 0.95" long slit and an integration time of 60 s at an airmass of 1.23. The 2100 lines/mm grating was used, providing a dispersion of 0.15 \AA / pixel. The spectrum was reduced, wavelength and flux calibrated using standard routines in IRAF and can be seen overlaid in Fig. \ref{fig:nebspec} .


\section{Analysis}

\subsection{Distance}

A distance estimate is given in  \citet{V906Car_Aydi_Nature_2020NatAs...4..776A} based on the UVES data set with Galactic reddening maps \citep{Chen_etal_2019} and a reddening value $E(B-V)$ = 0.36$\pm$ 0.05 mag, to derive a distance of $4.0 \pm 0.5$\,kpc for the nova.

For additional constraints on the distance, we carried out Gaussian fitting of the interstellar Na and K lines. The interstellar line profiles consist of 4 components in 
\ion{Na}{i} D and \ion{K}{i} 7699{\AA} lines and 6 in 
\ion{Ca}{ii} 3933
\& 3968{\AA} lines, see Fig \ref{fig:nebspecdist}, which indicates that the line-of-sight absorption consists of a number of discrete clouds each of which has some specific mean radial velocity with respect to the local standard of rest ($V_{\mathrm{LSR}}$). 
In particular, velocities derived from both \ion{Na}{i} D1, D2 and \ion{K}{i} lines indicate a range of mean cloud velocities of $-37$ to $-7.5$\,km\,s$^{-1}$ whereas the 
\ion{Ca}{ii} lines indicate $-43$ to $-1$\,km\,s$^{-1}$ range. If the velocity is due to Galactic rotation it can be used to derive a lower limit to the distance. In particular, the fourth component in 
\ion{Na}{i} D and \ion{K}{i} indicates a distance 
$d \gtrsim$ 4 kpc whereas the fifth component in \ion{Ca}{ii} indicates distance $d \gtrsim$ 4.5 kpc.  

The velocity and intensity structure of the interstellar \ion{Na}{i} D and \ion{Ca}{ii} lines in V906~Car is practically identical with those in two nearby well studied B1 Ib stars, HD 93827 and HD 94493 (see fig. 4 of \cite{1993A&AS..100..107S}). While their line of sight are very close to each other, ($l, b$) = (288.55, $-$1.54) and (289.01, $-$1.18) for HD 93827 and HD 94493, respectively, as well as to that of V906 Car (286.6, $-$1.09) the distance to HD 93827, $d$=8.3 kpc, is much larger than that for HD 94493 ($d$=3.3 kpc) which means that the interstellar lines can offer limited constraints for distances larger than $\sim$ 4kpc in the field/line of sight of V906~Car.
The reddening towards these two B1 Ib stars is also similar, $E(B-V)$=0.23 and 0.20 for HD 93827 and HD 94493, respectively \citep{1993A&AS..100..107S}, consistent with a lack of interstellar clouds beyond a distance of $\sim$ 4\,kpc in this line of sight.

\subsection{Unblended line evolution analysis and modelling: Morpho-kinematics with $\textsc{SHAPE}$}

The asymmetry in the lines at earlier times is opposite to those of the X-ray lines reported in \cite{V906Car_xray_2020MNRAS.497.2569S}, i.e. X-ray region is bright in the blue, but in the optical regime spectral lines are brighter on their red side for coincident observations. Implying the X-ray emitting region is at the shell exterior, probably due to the X-rays being heavily absorbed by the gas and the dust compositing the main shell. 

Blending of spectral lines is common for all optical nova spectra, apart from very late time nebular spectra where the strong nebular [O~{\sc iii}] and [N~{\sc ii}] lines can often be orders of magnitude brighter than any remaining lines. The strongest lines of the Balmer series also remain visible during these late times, where H$\alpha$ may be blended with the [N~{\sc ii}] lines on either side.

However, during the early times, when the object is still bright enough to acquire observations of high signal-to-noise ratio (SNR). With many Fe~{\sc ii} lines in these spectra line blending is a problem. This is especially true if the brightest spectral lines are used for line modelling, as they are more often than not blended. This is true for even H$\alpha$, H$\beta$, [O~{\sc iii}] 5007,4958{\AA} and [N~{\sc ii}] straddling H$\alpha$. Therefore, care must be taken when choosing which lines to use to model the ejecta geometry. 

Telescope systems such as the eMerlin network, ALMA, large ground-based telescopes (with M1 > 8m diameter) and HST can begin to visually untangle nearby (less than 1\,kpc), fast ($V_\mathrm{exp} > 1000$\,km\,s$^{-1}$) nova shells in the first 100 days of the nova shell evolution, however this type of data product is not commonly available. Additionally, imaging data needs to be complemented by polarimetric and/or spectroscopic data to fully untangle these early time nova shells - as line blending can also be a problem for narrow-band imaging, especially for the faster expanding nova shell population. The V906 Car nova shell did not settle into its frozen nebular state until > 1 year following maximum, and even at a distance $\sim$4 kpc and an expansion velocity $\sim$500 km s$^{-1}$ it could be spatially resolved by the aforementioned space based and large telescopes.

Through visual inspection of the day 395 post-max UVES spectrum, several lines were marked to be most likely unaffected by neighbouring lines, see Fig. \ref{fig:unblended1b}.
The selection of isolated lines with sufficient SNR were identified.

The isolated med/high SNR lines that can be compared in this way are:
3444{\AA} (O~{\sc iii}), 3797{\AA} (H~{\sc i}), 3995{\AA} (N~{\sc ii}), 4686{\AA} (He~{\sc ii}), 5876{\AA} (HeI), 6087{\AA} (N~{\sc ii}), 6678{\AA} (He~{\sc i}), 7065{\AA} (He~{\sc i}), as well as 8599{\AA}, 8750{\AA}, 8861{\AA} of the Paschen series.

The He~{\sc ii} 4686{\AA} line had the highest SNR of the sample, as well as most level blue- and red-shifted peaks, such that it was chosen as the most appropriate line to perform $\textsc{SHAPE}$ fitting of. These lines were also unblended in the SOAR spectrum, also shown in Fig. \ref{fig:nebspec}, giving increased confidence in their identification. The main differences seen between the +395 and +1001 spectra is the cooling of the expanding shell, characterised by the growth in strength of [OIII] lines.

Nova spectral lines, once frozen into their nebular state, can reveal the inclination of the nova system as for any one geometry and axial ratio, although one or two solutions may exist, see \cite{2newshells}. Unfortunately, we do not know the structure of a nova shell before it has been resolved. The most one can do is estimate the shell geometry based on the resolved nova shell population. 

\begin{landscape}
\thispagestyle{mylandscape}
 \begin{figure}
\centering
\includegraphics[width=24cm]{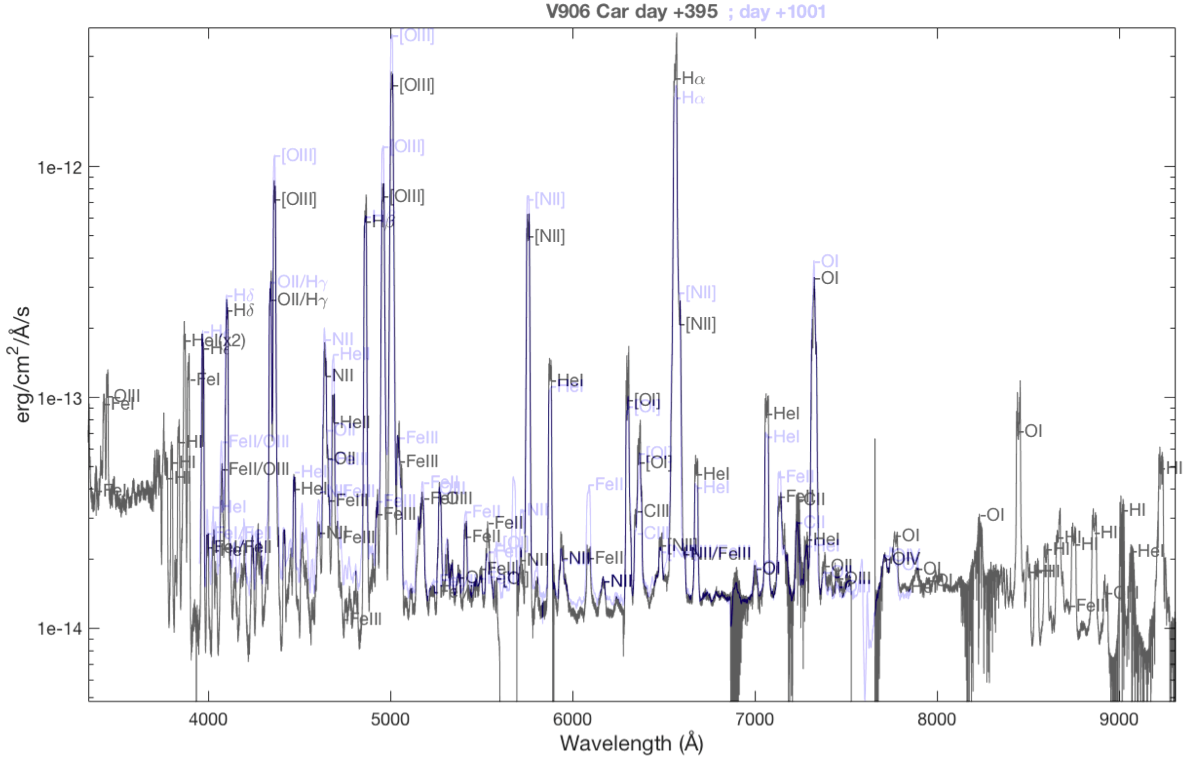}
\caption{UVES day +395 spectrum with a SOAR day +1001 spectrum overlaid in light blue. }
\label{fig:nebspec}
\end{figure}
\end{landscape}

\begin{figure*}
\centering
\includegraphics[width=15cm]{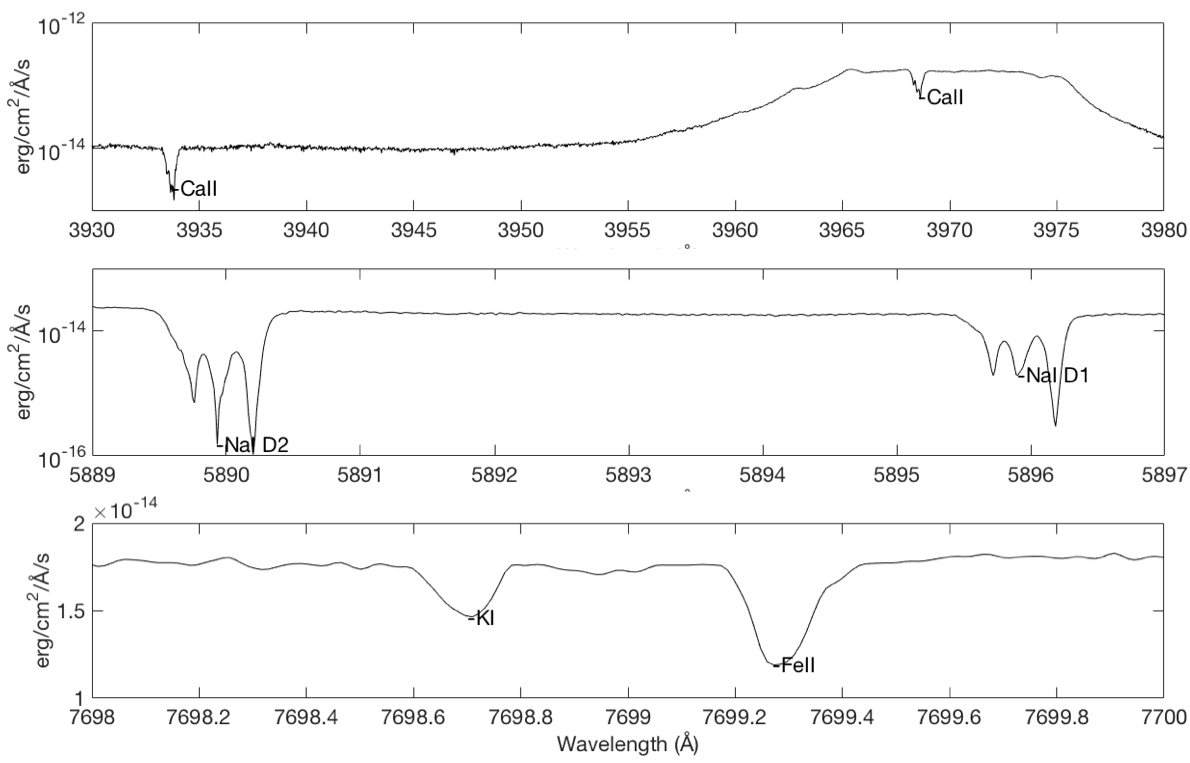}
\caption{UVES spectrum showing absorption components used for distance derivation in \citet{V906Car_Aydi_Nature_2020NatAs...4..776A}.}
\label{fig:nebspecdist}
\end{figure*}

\begin{figure}
\centering
\includegraphics[width=9cm]{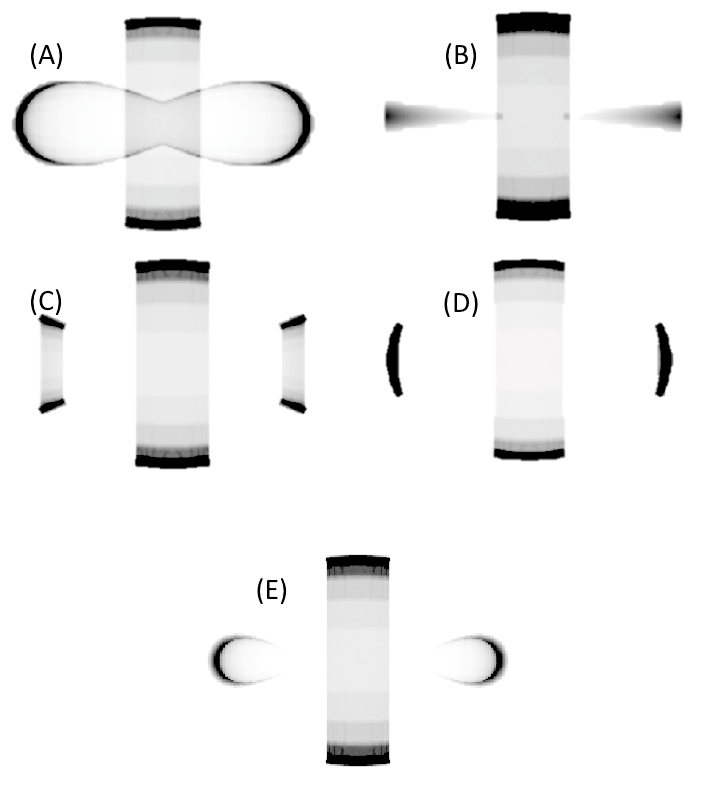}
\caption{Options for bipolar shell with equatorial waists. Best fit inclinations were derived for type (A) through to (E) assuming an axial ratio from 1.2 - 1.8. From these options Type (E) with an axial ratio of 1.4 was chosen for subsequent investigation.}

\label{fig:polarshapesb}
\end{figure}

\begin{figure}
\centering
\includegraphics[width=9cm]{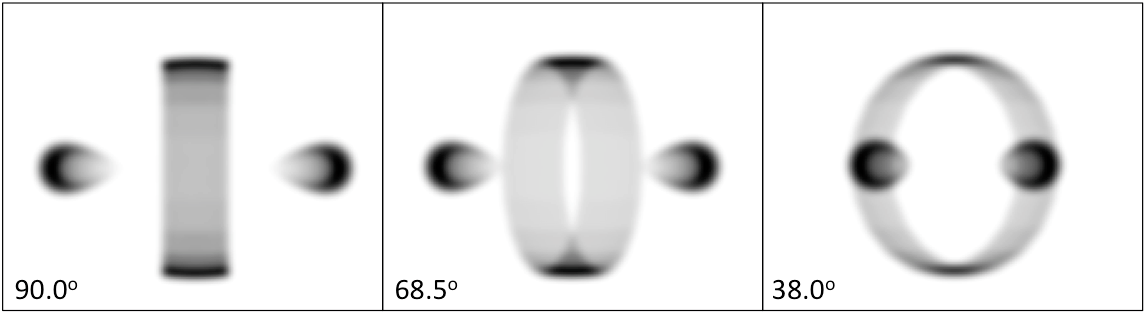}
\caption{Type (E) shell of Fig. \ref{fig:polarshapesb} for fitting geometry at 90$^{\textrm{o}}$ and the best fit inclinations for the observed spectral lines, i.e. 68.5$^{\textrm{o}}$ and 38.0$^{\textrm{o}}$.}

\label{fig:polarshapes}
\end{figure}

\begin{figure}
\includegraphics[width=9cm]{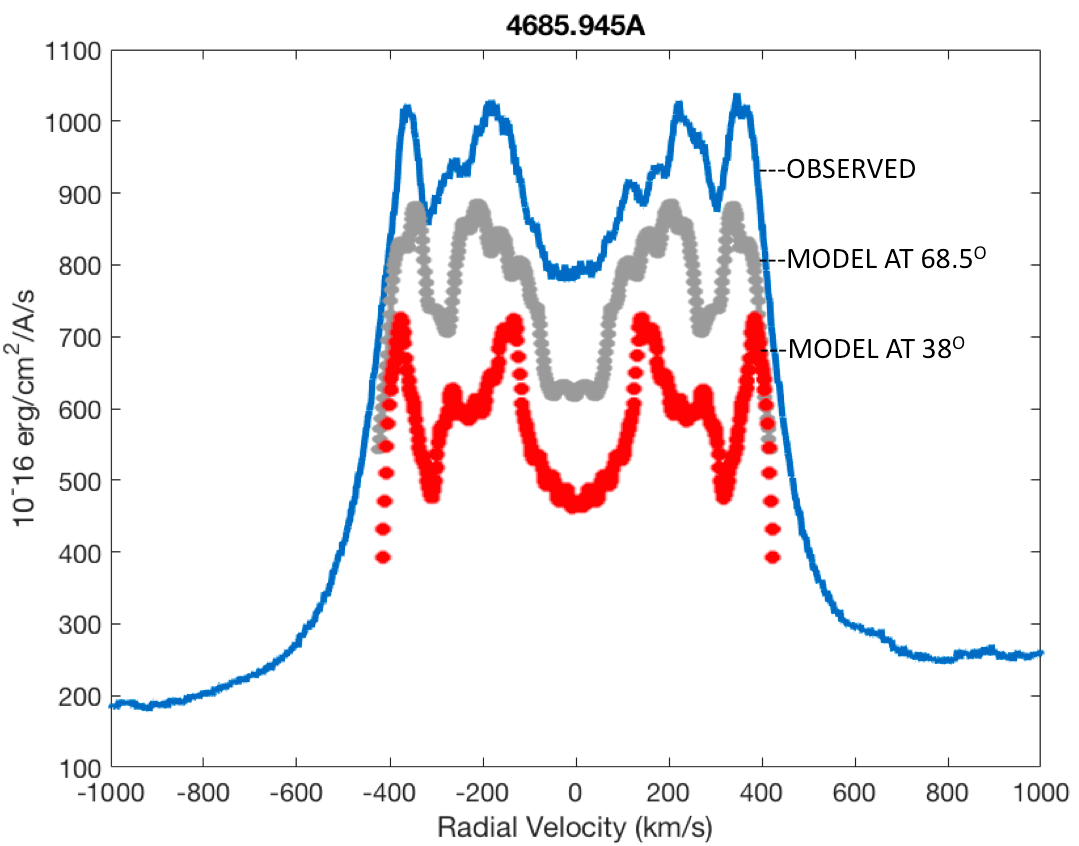}
\caption{He~{\sc ii} fit of spectral line 'crown'.  Grey is 
68.5$\protect{^{\textrm{o}}}$ and red is 38$\protect^{\textrm{o}}$ inclination and. Two solutions from geometrical model shown in Fig. \ref{fig:polarshapes}. The models have been offset for clarity.}
\label{fig:unblendedHeII}
\end{figure}

\begin{figure*}
\centering
\includegraphics[width=16cm]{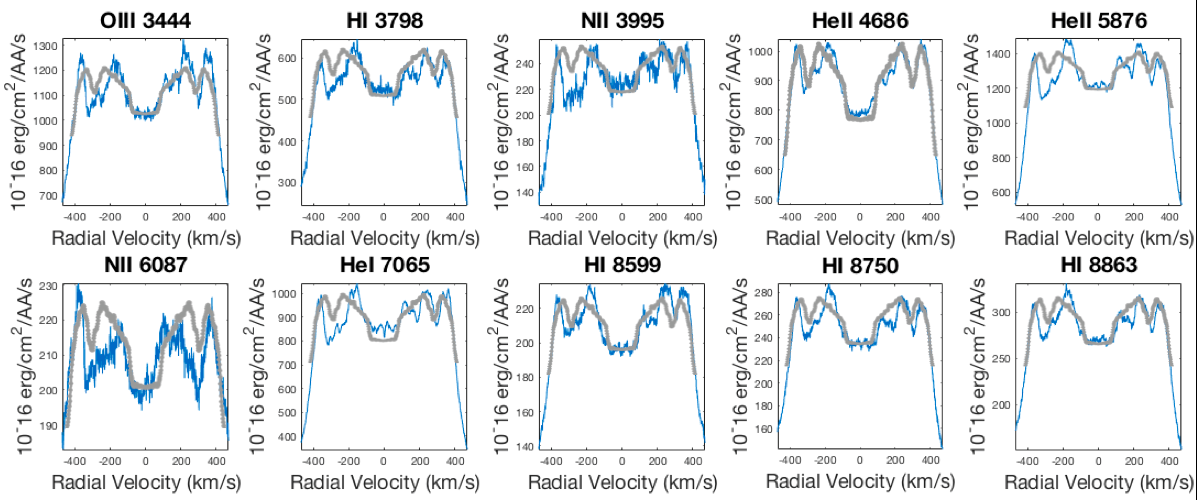}
\caption{Unblended lines with $\textsc{SHAPE}$ fit of at 68.5$^{\textrm{o}}$ inclination ($\textsc{SHAPE}$ model overlaid in grey), Figs. \ref{fig:polarshapes} and \ref{fig:unblendedHeII}}. 
\label{fig:unblended1b}
\end{figure*}

As the fitting geometry, a simple equatorial waist and polar feature morphology was used, as it is a commonly employed basic physical structure associated with nova shells, e.g. \cite{ontheapshericityofnovaremnants,Ribeiro2011,almaV5668Sgr,GKme,me_V5668Sgr,2newshells}, see Figs. \ref{fig:polarshapesb} and \ref{fig:polarshapes}. Morpho-kinematic modelling was undertaken using the SHAPE code \citep{shape}. Although, it is normal to observe tropical rings in nova shells also, they are left out of the model for the inclination fitting routine in order to have fewer starting elements. As they are intermediate between  equatorial and polar features their inclusion is less important in line shape fitting. Tropical rings are often present closer in azimuthal angle to the equatorial portion of the shell, as such they can have the affect of broadening the equatorial spectral components, especially for higher inclination systems. Without knowing the true deprojected velocity of the system inclination determinations remain first-pass. This could be solved with high-resolution multi-epoch narrow-band imaging of the resolved shell, or high-resolution IFU spectroscopy. 

The best fit inclinations given are for an axial ratio of 1.4, see Figs. \ref{fig:polarshapesb} and \ref{fig:polarshapes}. These being 38.0$^{\textrm{o}}$ and 68.5$^{\textrm{o}}$.
The approximation is based on that on average deprojected nova shells have axial ratios from 1.0 - 1.8, see for example Fig. 6 of \cite{Bode:2002aa}. For a different axial ratio the velocity of the polar  versus equatorial features  would vary by the same ratio and this would give different inclination solutions.
Then for an axial ratio of 1.0 inclination solutions of 32$^{\textrm{o}}$ and 60$^{\textrm{o}}$ are found. Additionally, for an axial ratio of 1.8 system inclination solutions of 44$^{\textrm{o}}$ and 76$^{\textrm{o}}$ are found. The $\textsc{SHAPE}$ model fits apply only to the top of the emission line since the line is further broadened by the shell sitting on higher velocity pedestals, i.e. lower density, higher velocity outflows also associated with the nova event \citep[again see][]{V906Car_Aydi_Nature_2020NatAs...4..776A,2020ApJ...905...62A}.


As seen earlier, two inclinations are possible with this given geometry of polar caps and equatorial ring, i.e., 38$^{\textrm{o}}$ and 68.5$^{\textrm{o}}$. This is because it is not known if the equatorial waist, or the polar cones are contributing as the faster component in the 1D spectral line, as this is inclination dependent. Interestingly, the average of our estimates is the same as that of \cite{2020MNRAS.495.2075P}. Figure \ref{fig:unblendedHeII} shows both fits, i.e. whether the equatorial waist is contributing to the faster peaks of the lines (68.5$^{\textrm{o}}$, dashed grey fit), due to line of sight effect, or whether higher velocity peaks are associated with the polar cones are responsible for the observed higher velocity components (38$^{\textrm{o}}$, dashed red fit). 
The almost flat section of the 68.5$^{\textrm{o}}$ 
fit between -100 and 100 km s$^{-1}$ is due to an averaging out of the plane-of-the-sky movement of the polar and equatorial components, i.e. a mix or a projection effect and assumptions on geometry. 

The fit in Fig. \ref{fig:unblended1b} is discussed here in an attempt to extrapolate the shaping history of the shell given what we now know from  \cite{V906Car_Aydi_Nature_2020NatAs...4..776A}, where an unequivocal relationship was observed between the $\gamma$-ray and optical emission from the early evolving nova event. As the shocks most likely originate from the equatorial region \citep[at least initially;][]{metzger}, this lends itself as a likely explanation, implied by viewing angle effects on the radial velocity measurements. For later shocks one might expect them to consist of two components, an equatorial shock followed by shocks at the poles between the swept up material from the previous interactions as well as the most recent, at a given time.

\begin{figure*}
\centering
\includegraphics[width=18.5cm]{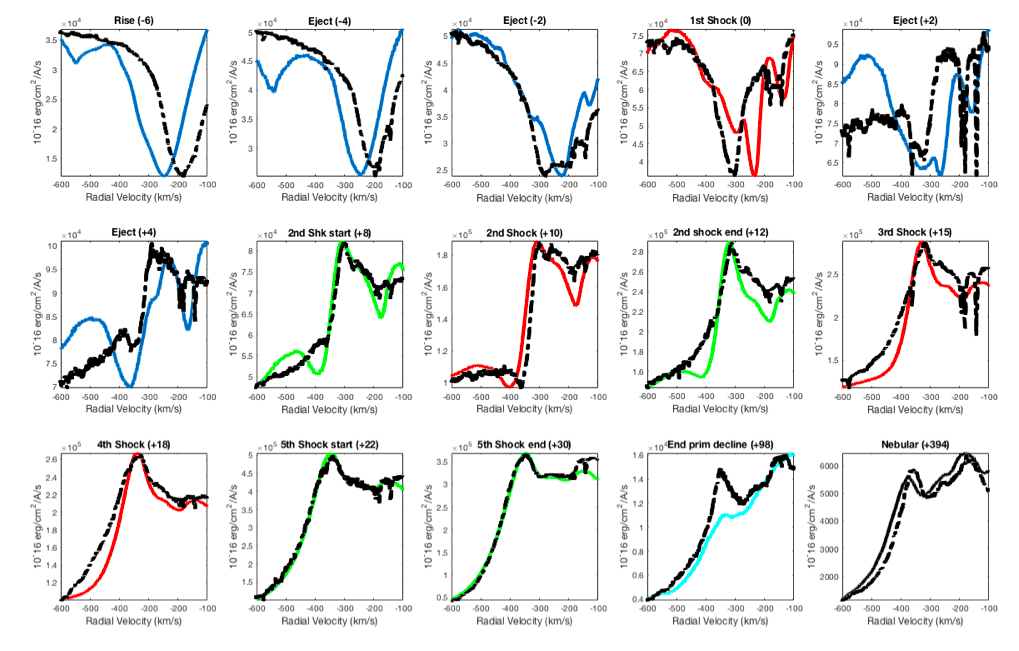}
\caption{Evolution of H$\beta$ (solid colour lines) in comparison to the 8863 {\AA} Paschen series line (black dot-dashed overlay) at corresponding epochs. To highlight a particular interpretation of the optical spectra and light curve the colour coding of the H$\beta$ follows blue for optical rises and peaks not thought to be associated with shocks, red for spectra taken during suspected shocks, green for pre- and post-shock spectra, cyan for the end of the primary decline and black for nebular. As elsewhere in this paper day 0 is defined as when the first suspected shock took place. Such that, for example, the second panel, labelled (Eject (-4)), corresponds to a suspected ejection event four days before the first suspected shock. }
\label{fig:HbetaPas}
\end{figure*}

Interpretation of  \cite{V906Car_Aydi_Nature_2020NatAs...4..776A} allows to associate UVES spectroscopy observation dates with the following nova phenomenology, as can be seen in Figure \ref{fig:HbetaPas} : \newline
\textit{Max -4}: Follows the first probable ejection. Average P-Cygni radial velocity: -200 km s$^{-1}$ (measured from \ion{Na}{i} D 5889.95 A line). The initial shell ejection has begun, understood to be more massive and slower than the subsequent ejection episodes. For now the P-Cygni line profile is best matched by a blue inverted fit. The blue absorption dip is characteristically narrow, due to it coming from a single, finite ejection. The velocity matches that of the inner peaks of the nebular lines - this is thought to be coincidental. However, the inverted peak fit allows for the visualisation of the acceleration of the main shell in Fig. \ref{fig:HbetaPas}. 
The red emission component is not yet visible. 
\newline
\textit{Max -2}: During second ejection. Average P-Cygni radial velocity: two components: -200 km s$^{-1}$ and -280 km s$^{-1}$. The blue-shifted absorption profile broadens and increases in velocity. This is believed to be due to the pre-shock scenario where the slow flow visible at Max -4 and a new higher velocity, less massive absorption component are unresolved from each other. 
\newline
\textit{Max 0}: During first hypothesized shock. Average P-Cygni radial velocity: two components: -220 km s$^{-1}$ -300 km s$^{-1}$. An increase in blue absorption line velocity, this is expected to be from the interaction of the line as described for Max -2 and a 600 km s$^{-1}$. Outflow traced in the Na D lines by \cite{V906Car_Aydi_Nature_2020NatAs...4..776A}. Now the red shifted emission component is beginning to appear in the Paschen series. The blue profile is beginning to $\lq$grow' out of its absorption trough, with internal reddening effects becoming less important as the shell expands. \newline 
\textit{Max +4}: After 3$^{rd}$, 4$^{th}$ ejections and during 5$^{th}$  ejection. The entire line is now in emission, however due to optical depth effects the blue component is still partially absorbed. \newline 
\textit{Max +22}: Follows Fermi-LAT's second detection, i.e.,  post-shock. Fermi-LAT's two strongest detections show a secondary lower amplitude episode. This can be explained by the initial stronger component arising from interactions at equatorial latitudes, followed by shocks in the system's polar direction. This implies that Fermi-LAT did miss an earlier shock episode during its downtime. Fig. \ref{fig:HbetaPas} demonstrates a change between Max +4 to Max +22 to an averaging of Max +4's two blue-shifted absorption components, signifying the meeting of two ejection episodes. \newline 
\textit{Max +394}: Now the shell is in the nebular stage, i.e., all major ejection and shock phenomenology have ceased and the forbidden lines are strong. The thinned and shaped ejecta has now frozen in structure, although continuing in expansion. Final average shell radial velocity is 370 km s$^{-1}$.  All previous absorption components have become their emission counterparts. Those that were due to later ejections that ran into the earlier ejected bulk shell, are now seen as the pedestal on which the line sits. \newline

Here we examine only the lower velocity features to look at the nova shell growth. For a summary of the high velocity features see \citet{V906Car_Aydi_Nature_2020NatAs...4..776A}. The time of appearance and velocity of the ejecta were traced in the Na D lines in \cite{V906Car_Aydi_Nature_2020NatAs...4..776A}. Three ejecta velocities from \cite{V906Car_Aydi_Nature_2020NatAs...4..776A} are recorded as $v_{1}$ < 600 km s$^{-1}$, $v_{2}$ $\sim$ 1200 km s$^{-1}$ and $v_{3}$ $\sim$ 2500  km s$^{-1}$.
The post-shocked components of the higher velocity features are eventually seen in the final nebular shell lines as the pedestals of which the crown fits sit on, refer to H\,{\sc i} 8863\,{\AA}.

In the context of the 38$^{\textrm{o}}$ model it could be an outflow with an equatorial over-density that would be seen to emerge first. Then, subsequent ejections crash into the initial outflow and get funnelled towards the poles, possibly creating tropical rings, akin to the $\lq$sweeping-up' as described in \cite{ontheapshericityofnovaremnants}. The polar material is expected to be relatively unimpeded following the first shock episode, subsequent shock episodes are then expected to have a different $\gamma$-ray light curve profile to account for both equatorial and soon later the polar material being shocked. This is expected to give double peaked gamma-ray light curves, however forward and reverse shocks may also give this effect. 

The central part of the line profile, which is stronger earlier on (and falls following the second hypothesised shock, i.e. between day +12 to  +15) may be associated with the pseudo-photosphere, accretion disk, a section of the equatorial disk, or even part of a more isotropic early outflow that is subsequently shaped. \cite{V906Car_xray_2020MNRAS.497.2569S} state that this nova experienced no distinct super soft phase, possibly due to a low mass white dwarf.

To uncover shock footprints it is best to uncover isolated lines, as well as the Na D absorption lines, from before, during and after the shock episodes. Each of the three Fermi-LAT detection episodes is followed by a more minor episode. This is possibly due to (a) an equatorial shock, followed by a shocks in the polar directions; or (b) the double peaks in the Fermi-LAT detections could be due to forward and reverse shock signature; or (c) shocks at different layers of the ejecta. Considering the hypothetical first shock event (which is thought to be missed during the Fermi-LAT down-time), if (a) is true, should only have a single peak, from a shock in the equatorial region. 

Later shocks should give the investigator a solution to the system axial ratio, provided the correct velocity components can be identified. Over the course of an individual shock episode the faster component should lose velocity on the average and diminish in strength. There should exist the two pre-shock velocities, one each for the slower and faster components, the appearance of a new shock velocity (equivalent to the difference between the two ejecta fronts and depending on relative densities), and a post-shock velocity.  Motivated by this we seek here to find the corresponding velocity components within the UVES spectral data set.  

As usual with novae, care must be taken with line overlap. The blended H$\beta$ line and the relatively unblended Paschen series line at 8863 {\AA} are used to demonstrate this exercise (Fig. \ref{fig:HbetaPas}). We find, initial outflow velocity of -200 km s$^{-1}$ and a later ejection velocity of -540 km s$^{-1}$. During the first (suspected) shock a transient velocity feature appears at -520 km s$^{-1}$. This transient velocity feature reduces to -460 km s$^{-1}$ after eight days, then two days later this feature jumps back up to -500 km s$^{-1}$ before being washed out. The arrival of this -500 km s$^{-1}$ component corresponds to the second suspected shock. For a summary of the higher velocity components observed in the UVES spectra see \cite{2020ApJ...905...62A}. A challenge encountered is in determining when specific line components should be in emission or absorption, for this we refer to Fig. \ref{fig:HbetaPas}. The optical depth of the region under study is dependent on the relative placements of the various components with respect to each other at that point in time. However, blending of even the strongest lines with weak neighbours can lead the investigator to derive incorrect expansion velocity values for the various components. 

In Fig. \ref{fig:HbetaPas} it is evident in the earlier spectra that the H$\beta$ line appears artificially faster than the Paschen series counterpart due to the proximity of a line on its blue side. Highlighting the need to explore unblended lines for this type of analysis. However, comparing unblended lines with strong blended lines lends information on what line features are real. This implies that the Paschen 8863 {\AA} line should be used to monitor the nova shell evolution for this nova. Referring back to Fig. \ref{fig:HbetaPas} and analysing the 8863 {\AA} line (dot-dash black line), a characteristic $\lq$step' can be seen where in the first panel a broad P-Cygni is observed, followed by a thinning of the P-Cygni profile in the second panel (labelled eject (-4)) and gradual appearance of two additional (narrow) superimposed absorption components. These then become more apparent in the third panel (Eject (-2)) where the line appears to broaden but instead is comprised of at least three distinguishable velocity components. Then in the fourth panel the P-Cygni profile has increased in velocity and has narrowed - the spectral line signature of a shock. 

Following this the relative densities of the separate outflows can be determined if the two responsible velocity components can be disentangled. Interestingly, weak during this shock and stronger in the next two panels we observe two sharp and apparently lower velocity absorption spikes. With the suspected equatorial (side-on) view of the system these $\lq$lower' velocity components may indeed be the deprojected higher velocity polar components forming. In the middle panel, the suspected second shock, these two lower radial velocity features weaken and by the panel marked the end of the suspected $2^{nd}$ shock when the the main equatorial component (now in emission) narrows again. This is followed by the lower radial velocity components becoming important once more - falling in line with the idea of excitation of the polar components. This would make sense in a system where following an equatorial shock, some material is swept up or $\lq$funnelled' in the polar direction. In this picture the shocks are shaping the shell. In later shocks, as the polar components should be well formed according to the proposed mechanism, an equatorial shock should then be followed by a polar shock. However, a means to separate out forward and reverse shocked components as well as equatorial and polar shocks would require similar observations at a higher cadence with coincident $\gamma$-ray and, ideally, radio observations.

\subsection{Density and temperature estimates}

\begin{figure*}
\centering
\includegraphics[width=17.5cm]{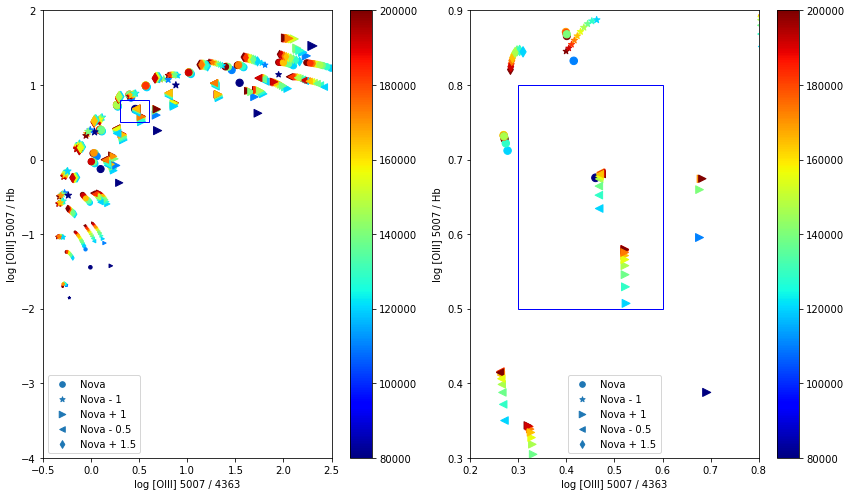}
\caption{Grid showing the models run for this work in an attempt to pin down the blackbody temperature of the ionising source. The size of the markers are related to the shell hydrogen density, the colours related to the blackbody temperature, the shapes to the shell abundances and deviation from in dex. The boxes indicated the measured values from our observations. We find that a range of abundances and temperatures can account for the observed line ratios. }
\label{fig:mormodgrid}
\end{figure*}

In order to refine the parameter space for deriving solutions to shell temperatures and densities we used the photoionisation code $\textsc{Cloudy}$ version 17.00 \citep{2017RMxAA..53..385F}, as well as several python wrappers available for it, namely $\textsc{pyCloudy}$ \cite{pycludy}, $\textsc{pyNeb}$ \cite{2015A&A...573A..42L} for diagnostics and, finally, $\textsc{pyCROSS}$ \citep{2020A&C....3200382F} that pairs $\textsc{SHAPE}$ and $\textsc{Cloudy}$. 
We begin by fixing the shell inner radius by assuming a bulk shell velocity of 450 km s$^{-1}$, giving $1.25\times10^{15}$ cm inner radius), distance (4kpc), filling factor (1\%) and log luminosity (4.2 L$_{\odot}$), i.e. from spectral fitting of the +395 day UVES observation and a sanity check to the Balmer line ratio matching to the \cite{2019MNRAS.483.4884M} model grid. Then, as free parameters we iterate over $T_{eff}$ (80,000 K to 200,000 K), nH ($10^6$ to $10^{9}$ cm$^{-3}$) and the logarithmic shift of the metallicity (-1 to +1 dex) relative to the $\textsc{Cloudy}$ "NOVA" range exists in the early onset. For this grid instead of using an outer radius we assume the nova shell is radiation bound. This is since greater velocities exist in the early spectra that are weaker in emission (the pedestals) on day +395. Instead we use a simulation stop condition of a temperature of 100 K. Results can be seen in Fig. \ref{fig:mormodgrid}, where the observed ratios are overplotted as boxes. Immediately problems are evident, solutions to the observations are applicable over a large range of temperatures and metallicities. As such T$_{eff}$ is not determined in our model grid as every T$_{eff}$ cycled over fits within the observed box of Fig. \ref{fig:mormodgrid}. We also find that log(U) is not really a constraint and, regardless, will not give a reliable value of nH given uncertainties in the radius, filling factor and luminosity of the ionising source. Interestingly, a broad set of abundances can replicate the observed results for the often used [OIII] diagnostics, i.e. the nova abundances as well as +1 dex and -0.5 dex changes to them. This analysis demonstrates the challenges behind the curtain of using single effective temperature estimates for single electron density solutions, when in general a family of solutions exist within the nova shell parameter space, sitting between the neat power law dependencies of lower density and lower energy planetary nebulae and the higher energetics of supernova remnants.

\begin{figure}
\centering
\includegraphics[width=8.5cm]{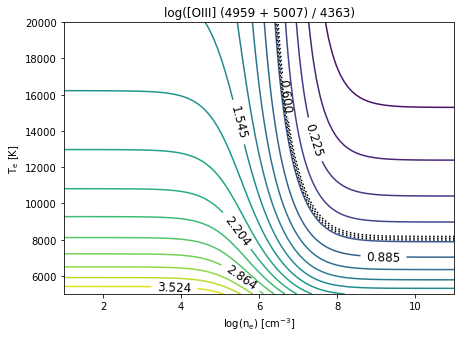}
\caption{[OIII] diagnostic, with observed ratio marked by the dashed line. This demonstrates that the commonly used temperature diagnostic line ratio is instead an electron density ratio for that observed here.  }
\label{fig:oiiidiag}
\end{figure}

\begin{figure}
\centering
\includegraphics[width=9cm]{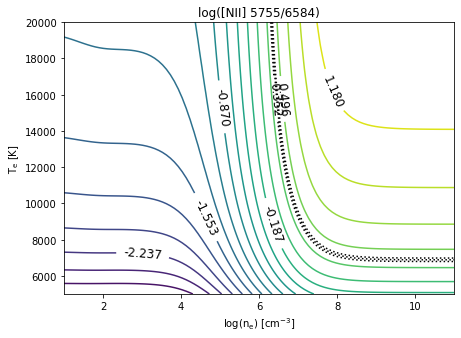}
\caption{[NII] diagnostic, again with the dashed line showing our measured line ratio. Similarly to the [OIII] diagnostic we have a strong lower bound of the electron density.}
\label{fig:niidiag}
\end{figure}

When considering the reddening correction there is an issue where some claim EBV to be 0.75 \citep{2020MNRAS.495.2075P}. However, from our observations (i.e. the same used in \cite{V906Car_Aydi_Nature_2020NatAs...4..776A}) the reddening is of the order of 0.35. This latter value would give H$\alpha$/H$\beta$ $\sim$ 2.85, which suggests an electron density of $<$ $1\times10^{8}$ cm$^{-3}$. On investigation of Figs. \ref{fig:oiiidiag} and \ref{fig:niidiag}, products of the model grids as detailed at the beginning of this section, we get a lower limit to n$_{e}$ of $\sim$ $5\times10^{6}$ cm$^{-3}$. 
We took measurements of the [NII] and [OIII] line ratio diagnostics from the line central plateaus and the secondary red shifted line peaks. The [NII] diagnostic gave no discernible differentiation, however the [OIII] diagnostic indicates that for any given T$_{eff}$ in the range a corresponding lower density is found for the central plateau than the red peak, and/or a lower temperature for a given density. 


Looking at Figs. \ref{fig:oiiidiag} and \ref{fig:niidiag}, which were created using pyNeb \cite{2015A&A...573A..42L}, we see that the typical [OIII] and [NII] diagnostics paired with the reddening would suggest likely electron density in the region of $(1-5) x 10^7$ cm$^{-3}$. However, the temperature is poorly constrained. 

Looking at the largest radius ($10^{15}$cm) and thickest shell (also $10^{15}$cm) covered in the results tables of \cite{2019MNRAS.483.4884M} an n(H) of $\leq 10^{8}$ at a blackbody temperature of $(1-5) x 10^5$ K and bolometric luminosity of $10^{37}$ - $10^{38}$ ergs s$^{-1}$ are found.

\begin{figure*}
\centering
\includegraphics[width=13cm]{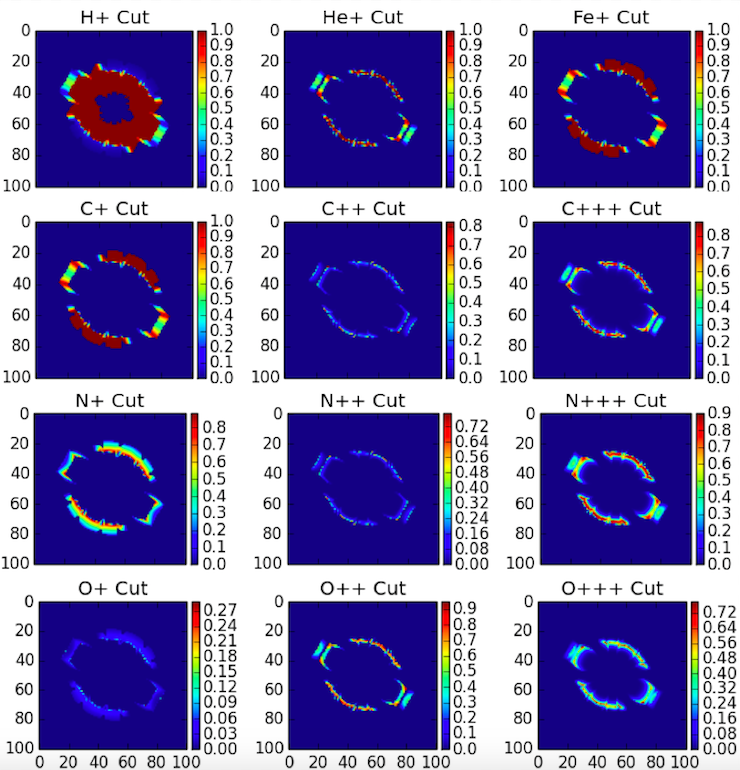}
\caption{$\textsc{pyCross}$ visualisation of various ionisation states of an equatorial structure and polar blobs (the components visible at top left and bottom right of each subplot) at 68.5$^{\textrm{o}}$. This grid shows the expected location of neutral H, He, C, N, O and Fe as well as the first and second ionisation states of C, N and O. The x and y axes are normalised distances on the plane of the sky. The colourbar represents the ionised fraction within the species. These simulations are for a purely photoionised nebula. If partially or wholly shock ionised then the ionisation stratification would be expected to deviate from that shown here. }
\label{fig:pycrosss}
\end{figure*}

Despite the difficulties in constraining the system, we wanted to simulate the ionisation structure of a purely photoionised look at the V906 Car shell at about a year post maximum. We used the above diagnostic information, including morphology, kinematics, temperature and densities, and with $\textsc{pyCross}$ \citep{2020A&C....3200382F} at hand we model the ionisation structure of the nova shell, see Figure \ref{fig:pycrosss}. Here, in this model neutral H is seen to occupy the entire shell, with neutral C and Fe brightest in the equatorial region. For lines associated with the Balmer series, they appear almost everywhere in the nebula. In our simulations He II lines, such as 4686\AA   
 are inner to He I lines, with He II near identical to the C++ and N++ panels in Figure \ref{fig:pycrosss}. Whereas [OIII] appears almost everywhere, showing that the observable nebula is cooling as a whole. O  I lines are the most localised, similar origin to He II but over a more narrow area. These ionisation structure models can be used to further interpret the spectral line profile evolution of the shell. However, these models show photoionisation stratification, where a different stratified structure would be expected to be observed if shocks contribute to the ionisation budget in part or in whole.

$\textsc{Cloudy}$, as used for the parameter search, assumes spherical symmetry or an open geometry (i.e. a small opening angle). Using 
$\textsc{pyCross}$ it is possible to run 
$\textsc{Cloudy}$ over axi-symmetric geometries, such as bipolar nebulae, it also hosts the ability to define different density laws for various regions of the nebula. However, here this was not done as we did not have the diagnostic information to separate components. Although possible with high resolution line spectra, this would be better done with IFU spectroscopy of a spatially resolved nebula. One reason why we did show our attempt here is because we found the diagnostic lines used were not sufficient in their accuracy to disentangle the subtle density variations between shell components as their effect was drowned out when considering sources of error. In any case, we are limited to axially symmetric structures with $\textsc{pyCross}$. If the nova shell were to deviate from axial-symmetry, then our simulations would be inadequate as they do not cover the complexities that would imply. However, near symmetric structure of the unblended spectral lines during nebular stage, shown in Fig. \ref{fig:unblended1b}, suggest an axi-symmetric shell. For reference we show readily identified non Balmer lines and equivalent widths in 
Table. \ref{line_det_3}, which with 
Table. \ref{tab:lineratios} makes clear the measurable lines for the epoch under study.

In Table. \ref{tab:lineratios} we show a model fit from an electron shell density of 7.5 dex. 
The source distance was set at 4 kpc, covering factor of 0.3, filling factor of 0.01, radius $1.25\times10^{15}$ cm, luminosity solar $10^{4.2}$ (in agreement with \cite{2020ApJ...899..162W}), age 395 days and Cloudy V1500 Cyg nova abundances from \cite{1978ApJ...226..172F}. This model gives an electron temperature of 6,500 K, and an energy density temperature of 3600 K, such that grains would survive in the environment. In this context for O, photoionisation of upper levels of [O~{\sc i}]  reached 7\% of the total O destruction rate. Photoionisation of upper levels of [O~{\sc iii}] reached 4\% of the total O++ photoionisation rate.
Non-collisional excitation of [O~{\sc iii}] 4363 {\AA}  
reached 1\% of the total. This is the same fit as used to make Fig. \ref{fig:pycrosss}. This model is shown for reference only, given that the conditions are difficult to pin down given the uncertainties demonstrated in Figs. \ref{fig:mormodgrid}, \ref{fig:oiiidiag} and \ref{fig:niidiag}. The consequences of shocks can be ignored at this stage as the shell has entered the nebular phase, although as discussed in \cite{derdzinski} early shocks influence the expected shell density and clumping efficiency. 

Intrinsic and emergent line intensities for 1000+ lines for all grid models are available as supplementary data available online. Column depths and averaged quantities of ionisation parameters are also provided within the same data files, see Data Availability statement.

\begin{table}
\centering
\caption{Line ratio models and observations for H$\beta$ and the better known lines shown in Fig \ref{fig:unblended1b}. Where lines could not be traced in the model, lines of the same species are chosen as replacements. Columns represent line i.d., wavelength,  modelled and observed line ratios with reference to H$\beta$. We assume Case B recombination for H.}
\label{tab:lineratios}
\begin{tabular}{llllll}
\hline
line & $\lambda$ (\AA)  & Model (L/LH$\beta$) & Obs (L/LH$\beta$) &  \\ \hline
H    I & 3798  & 0.06 & 0.04 &  \\
$[$O~{\sc iii}$]$ & 4363  & 0.81 & 1.09 &  \\
He I & 4471 &  0.08 & 0.07 &  \\
He II & 4686  & 0.09 & 0.13 &  \\
H    I & 4861  & 1 & 1 &  \\
$[$O~{\sc iii}$]$ & 5007  & 2.56 & 2.68 &  \\
He I & 5876  & 0.29 & 0.21 &  \\
He I & 7065  & 0.19 & 0.15 &  \\
H I & 8598  & 0.02 & 0.03 &  \\
H I  & 8750  & 0.02 & 0.03 &  \\
H I & 8863  & 0.02 & 0.04 &  \\ \hline
\end{tabular}
\end{table}

\begin{table}
\centering
\caption{Log10 Column density ($cm^{-2}$). Fe(a) are $\textsc{Cloudy}$ nova abundance \citep{1978ApJ...226..172F}, which are based on the very fast nova V1500 Cyg \citep{1978ApJ...226..172F}. Fe(b) = -3.77, which is in reference to the work presented here see Table.
\protect{\ref{tab:Feiii}}.}
\label{tab:coldens}
\tabcolsep=0.11cm
\begin{tabular}{lllllllll}
\toprule
level & 0 & 1 & 2 & 3 & 4 & 5 & 6   \\
\hline
H & 22.424 & 21.461 & 14.987 &  &  &  &    \\
He & 21.449 & 19.861 & 18.521 &  &  &  &  \\
Li & 9.38 & 13.778 & 10.79 & 9.801 &  &  &    \\
Be & 8.558 & 11.872 & 10.294 & 8.615 & 6.559 &  &    \\
B & 9.801 & 14.341 & 12.517 & 12.115 & 10.183 & 7.398 &    \\
C & 15.874 & 19.431 & 17.801 & 16.373 & 15.627 & 13.05 & 5.141   \\
N & 20.411 & 19.419 & 18.64 & 17.455 & 16.283 & 14.972 & 11.819   \\
O & 20.659 & 19.584 & 18.748 & 17.646 & 16.548 & 15.133 & 13.427   \\
Ca & 11.848 & 16.755 & 16.016 & 14.134 & 13.441 & 12.527 & 11.342   \\
Fe(a) & 14.01 & 18.347 & 16.411 & 16.32 & 15.198 & 14.869 & 14.053   \\
Fe(b) & 8.723 & 15.716 & 17.759 & 16.797 & 15.03 & 14.591 & 13.051   \\ \bottomrule
\end{tabular}
\end{table}

\subsection{Fe abundance}
\label{Feabund}

The early evolving environment of novae is complex, as there is a combination of shock- and photoionisation exciting the layered gas and dust shells. As such there exists a large abundance discrepancy especially for Fe as determined from X-ray and optical diagnostics, see \cite{2009ARep...53..605S}. As the degree of ionisation from shock contributions has been historically difficult to determine, with there being additional ways to hide Fe from X-ray observations \citep{1997ApJ...477..128A}, a possible way to look at the problem would be in understanding the correlation between temperatures derived from coincident X-ray and optical observations \citep{1999ASPC..163..153L}.

Here, we calculate the nebular stage abundance of [Fe III]. It is cautioned that the abundance calculations depend on density, for which we assume the best fit value from the previous section, i.e., of 7.5 dex. Line blending is taken into account, for example the 5412 {\AA} line could be either He II or [Fe~{\sc iii}], or a blend of both and as such is not included in the calculation. We refer to the solar abundance value of Fe/H from \cite{2010Ap&SS.328..179G} of $3.16 \times 10^{-5}$. 

\begin{table}
\centering
\caption{This table shows log(Fe/H) for a list of close and unblended [Fe~{\sc iii}] lines. Determinations are for the lines listed and an average at the end of the table, i.e., where -3.77 = $1.70 \times 10^{-04}$, or 5.4 $\times$ (Fe/H) solar.}
\label{tab:Feiii}
\begin{tabular}{lll}
\toprule
species & $\lambda$ ({\AA}) & log (Fe/H) \\ \hline
\protect{[Fe~{\sc iii}]} & 4659 & -4.050 \\
\protect{[Fe~{\sc iii}]} & 4702 & -3.750 \\
\protect{[Fe~{\sc iii}]} & 4734 & -3.725 \\
\protect{[Fe~{\sc iii}]} & 4755 & -3.675 \\
\protect{[Fe~{\sc iii}]} & 4931 & -3.650 \\
\protect{[Fe~{\sc iii}]} & mean & -3.77
\\
\bottomrule
\end{tabular}
\end{table}

Looking at Table. \ref{tab:coldens} the column density distribution of the Fe level is shifted towards higher levels for the higher Fe nova abundance for V906 Car, as found in Table. \ref{tab:Feiii}. 
The large discrepancy in Fe abundances found between X-ray (< 0.1 Fe/Fe$_\odot)$ \cite{V906Car_xray_2020MNRAS.497.2569S}) and optical methods, found here (i.e., 5.4 Fe/Fe$_\odot$), implies that Fe could be hidden in the X-ray spectra. This can be achieved by acceleration of dust grains behind shocks. This is proposed by \cite{2009ARep...53..605S} where they argue the freezing-out of heavy elements is efficient due to betatron acceleration of dust grains behind shocks.

\section{Conclusions}

A methodology for uncovering blended lines has been laid out. This can be achieved through fitting a line at the nebular stage to develop a template. Then using line lists and ionisation codes lines can be separated, provided a sufficient number of unblended lines for that species exists for regular nebular diagnostics. There is scope for developing this methodology into a future fitting routine. It is suggested in this work that it is best to wait until a nova has entered the nebular stage in order to begin diagnostics, and then one can work backwards to try and understand the evolution up to the point of lines freezing into their final structure. 

It was shown that the fitting of a simple 3D morpho-kinematic model to a 1D line profile crown can aid in deciphering the evolutionary history of unblended spectral lines as well as aiding in the detection of weak lines. 
Sensitive spectroscopic observations are key to uncovering the underlying physics of these enigmatic events. It is suggested in this work that the slope of the broadened pedestal on which the line profile crown sits contains information of the faster flows from the nova. This information is restricted to velocity, as there is no discernible shape to the pedestal. Therefore if we assume the pedestal to belong to an isotropically expanding spherical component then we can take the ~900 km $^{-1}$, otherwise we can deproject its velocity using the 38$^{\textrm{o}}$ or 68.5$^{\textrm{o}}$ inclination solutions. 

\cite{V906Car_circumFeO_2020MNRAS.494..743M} state that the $\lq$M' spectral line shape can only be reproduced by a rotating disk, however we find that the sloped $\lq$drop-off' is well explained by the existence of faster moving material of lower inherent flux and that the width of velocity components is well explained by a thicker shell or even a thin shell/disk/ring/barrel viewed at any non-zero inclination. The argument presented in \cite{V906Car_circumFeO_2020MNRAS.494..743M} against the `M' profile arising from the shell is that it is too narrow (of too low a velocity) to arise from the shell. However, the observed line velocity is well represented as a component of the final (frozen/nebular) unblended shell line velocity.

In \cite{2020ApJ...905...62A}, the initial outflow is concentrated towards the equator, this may be due to the oblate nature of the fast spinning underlying white dwarf (see \cite{ontheapshericityofnovaremnants}) and/or the location of where material is falling onto its surface. The subsequent ejections are expected to also originate in an equatorial dominated outflow, however they are swept up into conical regions as they interact with the slow, dense initial ejection. The shocks between these outflows are the suspected origin of the $\gamma$-ray emission \citep{V906Car_Aydi_Nature_2020NatAs...4..776A, metzger}. The shock velocities are expected to be the difference in velocity of the interacting regions. 

It was found that during this investigation that both fast recurrent and slow classical novae tend to inhabit a rapidly varying density and temperature diagnostic space of commonly used nebular diagnostic lines. Meaning that one should tread with caution when making temperature and density estimates in novae. A larger sample of diagnostic lines of many species across the electromagnetic spectrum should be used to better constrain densities, abundances, reddening, filling factor, geometry and ionisation factors.
It would be advisable for the community to extend the grid parameters, in terms of radii and resolution of models from \cite{2019MNRAS.483.4884M} as well as lines covered, this would allow for a concerted diagnostic effort.

 In the resolved nova shell population where narrow band images of H$\alpha$ +  $[NII]$ and $[OIII]$ are common the general ionisation stratification is similar to that seen here, see T Aur \citep{gallagher1980}, HR Del \citep{Duerbeck,hutchingHRDEL,HRDel3d}, DQ Her \citep{williamsdqher,Vaytet}, V1500 Cyg \citep{beckerv1500cyg,hutchingsmccall}, V476 Cyg \citep{Duerbeck}, FH Ser \citep{gillbrien}, CP Pup \citep{Duerbeck}, RR Pic \citep{gill}, and GK Per \citep{Liimets:2012aa,GKme}. Unfortunately high resolution IFU observations (or channel maps) of easily resolved nova shells are rare. An example of such though are the IFU observations of  \cite{HRDel3d}, which also includes He II 4686\AA, and the GK Per channel maps of \cite{lawrencefp}.  

We find an Fe abundance discrepancy between X-ray and optical observations, parallel to that found in molecular clouds. The discrepancy between Fe abundances, found from optical and X-ray studies detected should contain shock information. It is also suggested in this work that spatially resolved narrow-band imaging of nova shells at early times should reveal if ionisation stratification is due to shock- or photoionisation, or both.

Prolonged and structured (e.g. waves/bumps) gamma emission and/or non-thermal radio emission in a growing number of classical novae has been observed over recent years \citep{cheung_gammav5668sgr,2017NatAs...1..697L,2018ApJ...852..108F,2022MNRAS.515.3028B}. Supporting the observations are a series of theoretical works, e.g.  \cite{metzger,derdzinski,2020MNRAS.491.4232S,2022ApJ...939....1H}. Slow evolving dust producing novae, similar to V906 Car, have a tendency to feature the brightest (V1324 Sco \cite{2018ApJ...852..108F}) and longest duration (e.g. V5668 Sgr \cite{cheung_gammav5668sgr}) gamma (>100 MeV) episodes. Where some have argued for a white dwarf spin correlation to the gamma periodicity (ASASSN-
16ma \cite{2017NatAs...1..697L}), here we posit several forward and reverse shocks in both polar and equatorial nova shell regions to give the observed structured optical and gamma-ray light curve of V906 Car, similar to the theoretical work of \cite{2022ApJ...939....1H}, effectively leading to the shaping of the nova shell as observed in its nebular phase.  \cite{2020MNRAS.491.4232S} discuss varying velocity outflows over the course of a nova eruption leading to flares and complex velocity evolution of absorption and emission components of spectral lines, which is what we propose here being observed in V906 Car.

Shocks lead to clumping of shell material \citep{massflows}, and the clump centres give a safe haven for dust formation in an otherwise harsh environment \citep{joiner}. Lending itself linearly to infrared peaks (i.e. dust) following gamma peaks (i.e. shocks, e.g., V5668 Sgr  \cite{banerjeeV5668sgr,me_V5668Sgr} and V809 Cep \cite{2022MNRAS.515.3028B})
and strongly-shaped dust-forming novae, e.g. T Aur \citep{gallagher1980}, HR Del \citep{Duerbeck,hutchingHRDEL,HRDel3d}, FH Ser \citep{gillbrien} and DQ Her \citep{williamsdqher,Vaytet}. This leads us to conclude that shocks likely shape nova shells in both their equatorial and polar shell shapes (giving the shell covering factor) and the clumping within these regions (the filling factor).

\section*{Acknowledgements}

ÉJH and MJD receive funding from STFC under grant ST/V00087X/1.
CM acknowledges the support from UNAM/DGAPA/PAPIIT grant IN101220.
E.A. acknowledges support by NASA through the NASA Hubble Fellowship grant HST-HF2-51501.001-A awarded by the Space Telescope Science Institute, which is operated by the Association of Universities for Research in Astronomy, Inc., for NASA, under contract NAS5-26555. EA also acknowledges NSF award AST-1751874, NASA award 11-Fermi 80NSSC18K1746, and a Cottrell fellowship of the Research Corporation.

\section*{Data Availability}

 Reduced data and scripts to run moels are available upon reasonable request. Raw data is publicly available through the ESO  and SOAR science data archive facilities. 
 \\

\begin{table*}
\centering

\caption{Spectral line identification of the optical spectra for lines not previously mentioned in Tables. \ref{tab:lineratios} and \ref{tab:Feiii} . We list the EW and integrated flux of the lines for which an estimate was possible (some lines are not strong enough or blended and therefore accurate measurements for these parameters were not feasible). The listed measurements are for +395 days after maximum.}
\begin{tabular}{rrrrr}
\hline
\hline
Line & $\lambda_0$  & EW ($\lambda$) & Flux \\ 
 & (\AA) & (\AA) & ($10^{-13}$ erg\,cm$^{-2}$\,s$^{-1}$) \\
\hline

$[$Ne~{\sc v}$]$ & 3346 & $-17 \pm 2$ & 5.8 $\pm$ 0.5\\
 $[$Ne~{\sc v}$]$ & 3426 & $-22 \pm 2$ & 7.7 $\pm$ 0.5\\
O~{\sc iii} & 3444 & $-25 \pm 2$  & 9.0 $\pm$ 0.5 \\
$[$Ne~{\sc iii}$]$ & 3869 & $-150 \pm 30$  & 21.3 $\pm$ 3.0 \\
$[$Ne~{\sc iii}$]$ & 3968 & --  & -- \\
$[$Fe~{\sc xi}$]$ & 3978 & --  & -- \\
$[$Fe~{\sc v}$]$ & 4071 & --  & -- \\
He~{\sc i} & 4144 & -- & -- \\
C~{\sc iii} & 4187 & -- & -- \\
$[$Fe~{\sc ii}$]$ & 4244 & -- & -- \\
C~{\sc ii} & 4267 & -- & -- \\
$[$O~{\sc iii}$]$ & 4363 & -- & -- \\
He~{\sc i} & 4471 & --  & -- \\

N~{\sc v} & 4609 & --  & -- \\
N~{\sc iii} & 4638 & $ -250 \pm 40 $ & 30 $\pm 2$ \\
He~{\sc ii} & 4686 & $ -100 \pm 10 $ & 12 $\pm 1$ \\
He~{\sc i} & 4922 & -- & -- \\
$[$O~{\sc iii}$]$ & 4959 & $ -800 \pm 100 $ & 134 $\pm 10$ \\
$[$O~{\sc iii}$]$ & 5007 & $ -2500 \pm 200 $ & 422 $\pm 20$ \\
He~{\sc i} & 5048 & -- & -- \\
He~{\sc ii} & 5412 & -- & -- \\
$[$N~{\sc ii}$]$ & 5755 & $ -950 \pm 50 $ & 104 $\pm 10$ \\
C~{\sc iv} & 5805 & -- & -- \\
He~{\sc ii} & 5876 & $-180 \pm 20 $ & 22 $\pm 2$ \\
N~{\sc ii} & 5938 & -- & -- \\
$[$O~{\sc i}$]$ & 6300 & $-200 \pm 20 $ & 22 $\pm 2$ \\
$[$O~{\sc i}$]$ & 6364  & -- & -- \\

He~{\sc i} & 6678 & $-60 \pm 5 $ & 8.0 $\pm 0.5$ \\

He~{\sc i} & 7065 & $-125 \pm 5 $ & 17 $\pm 1 $ \\
$[$Ar~{\sc ii}$]$ & 6136? & --  & -- \\
$[$Ar~{\sc ii}$]$ & 6237? & --  & -- \\
$[$O~{\sc ii}$]$ & 7320/30 & $-500 \pm 50 $ & 70 $\pm 10 $ \\
\hline
\end{tabular}
\label{line_det_3}
\end{table*}




\bibliographystyle{mnras}
\bibliography{nova_refs} 







\bsp	
\label{lastpage}
\end{document}